\documentstyle[aas2pp4]{article}

\def\Msun{\ifmmode M_{\odot} \else $M_{\odot}$\fi}
\def\Lsun{\ifmmode L_{\odot} \else $L_{\odot}$\fi}
\def\eg{{\it e.g.,\ }}
\def\ie{{\it i.e.,\ }}
\def\etal{{et al.~}}

\slugcomment{Not to appear in Nonlearned J., 45.}

\lefthead{Chatzichristou}
\righthead{IR-Warm Seyfert Galaxies}

\begin{document}

\title{Multicolour Optical Imaging of IR-Warm Seyfert Galaxies.\\
    III. Surface Photometry: Light Profile Decomposition}

\author{Eleni T. Chatzichristou}
\affil{Leiden Observatory, P.O. Box 9513, 2300 RA Leiden, The Netherlands}

\affil{NASA/Goddard Space Flight Center, Code 681, Greenbelt, MD 20771}



\begin{abstract}
This paper is the third in a series, studying the optical properties of a
sample of mid-IR Warm Seyfert galaxies and of a control sample of mid-IR Cold
galaxies. The present paper is devoted to surface photometry.
We analyse the light distributions characterizing the galaxies outside the
central 2 kpc. The radial light profiles are decomposed, using two generalized
exponentials, in inner and outer components. Each is characterized by the 
profile shape, central surface brightness and scale length. We find that 
light is more centrally concentrated in Seyfert 1s, that also tend to lie in
earlier-type hosts than Seyfert 2s. Seyfert 1 and 2 bulges have similar shapes
but the former are characterized by larger central surface brightnesses and 
smaller scale lengths.  The three parameters characterizing the bulge 
component correlate with each other, within a limited range of bulge 
luminosities. Cold galaxies are disk-dominated systems, with complex 
morphologies. Their bulges are flatter and fainter compared to the Warm 
sample. The disk structural parameters span similar ranges for the three 
(sub)samples but with larger scatter. The parametrization of light profiles,
as described in this paper, shows that the three (sub)samples occupy different
loci in parameter space, that is suggestive of an evolutionary connection 
between them.
\end{abstract}




%

\section{Introduction}

In the present (third) paper we continue investigating the optical properties 
of a sample of 54 mid-IR Warm Seyferts selected from the sample of IR-warm 
IRAS sources of De Grijp \etal (\cite{grijp87} and \cite{grijp92}). Our 
control sample contains 16 mid-IR Cold IRAS galaxies, selected to span 
similar redshift and luminosity ranges as the Warm sample.
In \cite{paper1} (hereafter Paper I) we presented our optical imaging data.
In \cite{paper2} (hereafter Paper II) we discussed and intercompared the
optical properties of these samples, resulting from our aperture photometry
and searched for correlations with their IR properties. In the present third
paper we will present, analyse and discuss the results of our surface 
photometry, performed on most of our sample objects. 

This paper is organized as follows: in Section 2 we summarize our method
of azimuthal ellipse fitting, two-component decomposition of the projected
1-D light profiles and their parametrization. In Sections 3 and 4 we discuss
the various structural parameters characterizing the light distributions and
intercompare our Warm and Cold (sub)samples. Our conclusions are summarized
in Section 5. 

\section{Light Profile Decomposition}

\subsection{Isophotal Fitting}

Most of the available isophotal fitting procedures approximate the galaxian 
isophotes with ellipses at increasing radii and subsequently perform surface 
and aperture photometry within each ellipse. The basic idea is to sample the 
image at predefined radii (or rather semi-major axis lengths) along an 
elliptical path, so that the intensity is the same at all sampling points 
within the noise. The intensity distribution along the ellipse is fitted by 
weighted least-squares to an harmonic expression of the type
\begin{eqnarray*}
y = y_0 + A_1 \times sin{(PA)} + B_1 \times cos{(PA)} + \\
+ A_2 \times sin{(2 PA)} + B_2 \times cos{(2 PA)}
 \end{eqnarray*}

{\em PA} being the position angle; the harmonic amplitudes 
$A_{1},B_{1},A_{2},B_{2}$ parametrize errors in the fitting procedure. Once 
the best fit ellipse has been obtained, the residuals along this ellipse are 
parametrized as
\[
 A_n \times sin{(n PA)} + B_n \times cos{(n PA)}, n=3,4
\] 
These higher harmonic amplitudes $A_{n},B_{n}$ characterize the deviations of
a given ellipse from perfect isophotometry. The method is described by 
\cite{jedrzejewski87}.
 
Among available ellipse fitting algorithms we have utilized the WFGAL 
sub-package of the IRAF/STSDAS applications, which is based upon a combination 
of tasks that are improved and/or extended versions of the original routines 
within the ISOPHOTE subpackage. Our method is described in \cite{thesis} and
was applied to all of our objects for which either photometric information
was available or which possessed well-resolved morphologies.

\subsection{Radial Profile Parametrization}

\begin{figure}
\epsscale{0.8}
\plotone{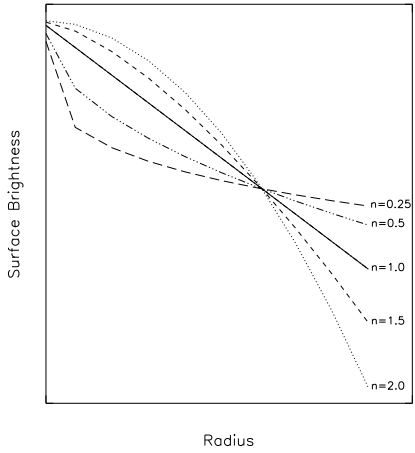}
\caption{The variation in shape of a Sersic (generalized exponential) profile as a function of the exponent value $n$. The radius where all profiles correspond to the same surface brightness level is the scale length $h$ (arbitrary). \label{f1}}
\end{figure}

\placefigure{f1}

Azimuthally averaged profiles have two important advantages over radial 
(usually along major and minor axes) profiles: first, they allow smoothing
of inhomogeneities that could be due to non-uniformly distributed dust, 
regions of enhanced star formation, or to the presence of non-axisymmetric 
features and second, they provide improved S/N ratio. 
The azimuthal averaging can be done without or after deprojection of the
galactic disk using an estimation of the galaxy's inclination {\em i}. Unless 
the galaxy shape is simple and well-defined, applying the latter method is 
subject to uncertainties due to the multiplicity of factors that can affect 
{\em i}, such as, projected dust absorption or real disk distortions.

The effects of seeing (due to Earth's atmosphere and scattering within the 
telescope) on the intensity and shape parameters of the fitted ellipses can be
very important and alter the results in the inner galaxy regions. Although our
analysis is mainly concerned with the extended component, we have applied an 
inner cut-off radius to our surface brightness profiles so that the profile 
decomposition (described below) is not affected by seeing effects.
Usually in such studies the minimum cut-off radius is taken to be half or 
equal to the seeing PSF, but a more elaborate method is described by 
\cite{franx89}. Using their formulae approximated for small core radii and 
taking \( \frac{\Delta I}{I} \) to be equivalent to the error in the local 
surface brightness due to the seeing effects, we have: 
\( \frac{\Delta I}{I}=\frac{F_2}{r^{2}} \)
where $F_{2}$ is the second order moment of the seeing PSF. For our data, we 
assume a ``reasonable'' value for \( F_{2}\simeq 0.8(FWHM)^{2} \) 
(see \eg \cite{franx89}; \cite{jorgensen92}. The FWHM for each object is 
measured on several stars of each image and the mean values are listed in 
Tables 4 and 5 of Paper I. Requiring that the error in the local surface 
brightness be less than 0.10 mag (\( \frac{\Delta I}{I}=0.1 \)) the inner 
cut-off radius is determined by \( r_{in}=\sqrt{8}\times(FWHM) \). Only in the
case of very compact objects, we reduced the cut-off radius to the value of 
the seeing disk. In any case, the choice of the inner cut-off radius in this 
study is not crucial, since we are not interested in the detailed structure of
the central regions of our galaxies.

Galaxy light profiles are in general more complex than a simple two component 
model, as indicated by a variety of observational studies. This is due to a
variety of factors:
(i) The structure (in particular at intermediate radii) can be very complex 
(\eg \cite{prieto92a} and \cite{prieto92b}), indicating the existence of one 
or more extra components: bars, rings, lenses (Freeman type II profiles),
spiral arms (\eg \cite{freeman77}; \cite{freeman70}; \cite{martin95}; 
\cite{serna97}). The presence of strong star formation in the inner disks 
mostly affects the $B$-band profiles (for small to intermediate redshift 
objects).
(ii) Dust extinction can affect seriously the observed light distributions,
especially at shorter wavelengths (\eg \cite{disney89}; \cite{valentijn90};
\cite{white92}; \cite{jansen94}).
(iii) Interactions with close companions are often responsible for deformed 
disks and tidal features, which increase the complexity of the projected light
profiles.
It is obvious that the use of any simple profile decomposition is bound to 
have limited success unless one accounts for all the above effects. It is 
thus often necessary to resort to a one-to-one comparison with the direct 
images and colour maps, to disentangle the effects of the various components.

\subsection*{\em The use of a Generalized Exponential Law}

Early studies have indeed shown that a simple bulge+disk profile decomposition
cannot describe properly more than about half of the observed spiral galaxies 
(\cite{boroson81}; \cite{kent86}). There is accumulated observational evidence
that the radial distributions of the spheroidal component of spiral galaxies 
(bulge) differ significantly from those of ellipticals (\eg\cite{kormendy78}; 
\cite{burstein79}; \cite{shaw89}; \cite{kent91}). Since some ellipticals and 
disk bulges are well-fitted by the De Vaucouleurs law but there are many that 
systematically deviate from it, the need for a generalized law with some 
additional parameter, became obvious. In fact, exponential and generalized 
exponential laws were successfully used to fit elliptical, S0 and disk galaxy 
bulge profiles (\eg \cite{kent91}; \cite{caon93}; \cite{andredakis94}
\cite{onofrio94}; \cite{jong96a}; \cite{jerjen97}). Most of these studies show
that exponential bulges are statistically at least as justified as De 
Vaucouleurs bulges. Independently of the model used for the bulges, an inner 
disk cut-off is required (\cite{kormendy77a}) to best fit the light profiles. 
This agrees with the observed gas distributions of many spiral galaxies. The 
cut-off could be due to the influence of the gravitational field of a central 
bar or the fall of material into the galaxy nucleus (\eg \cite{vaucoul70}). 
Whether bulge and disk components are physically distinct or not is unclear. 
\cite{caon93} \cite{andredakis95} and \cite{jerjen97} find a large range of 
values for the exponent {\em n} that describes the shape of the light 
profiles, consequently, the use of a single law for the description of the
profiles of galaxies with a variety of morphological types is clearly 
inadequate. This is especially true for galaxies with complex morphologies, as
is the case for most of our sample objects.

We attempted brightness profile decompositions based on a variety of bulge and
disk combinations. First, we defined the bulge and disk dominated 
regions on the brightness profiles, making also use of the ellipticity and 
position angle radial profiles: (i) The bulge dominated region should appear 
linear in the surface brightness vs r$^{1/4}$ plots, starting just outside the
radius where seeing effects are important. (ii) The disk-dominated region 
should be linear in the surface brightness vs radius plots,with $\epsilon$ and
{\em PA} remaining constant all the way out (until the profiles become too 
noisy or a tidal feature is present). The intermediate transition zone, 
present in most of our light profiles, was systematically excluded from the 
fits. We applied a $\chi^{2}$ polynomial fit to one of the two components, 
using (i) a De Vaucouleurs r$^{1/4}$ or an exponential law for the bulge and 
(ii) a simple exponential disk profile. We subtracted the best fit solution 
from the initial profile and fitted the residual with the second component. 
This best fit solution was then subtracted from the original profile. The 
whole process was repeated iteratively until convergence was achieved. For 
many of our profiles, we repeated the fits starting both from the inner 
(bulge) and the outer (disk) regions, to check the validity of the results and
the uncertainties in the fitted parameters. In general, these were similar for
both methods. The best fit results, obtained in this way, provided initial 
estimates for the spheroidal and disk components and were used to 
simultaneously fit the two components in the next step. The composite 
bulge/disk fitting was done over the whole profile range for relatively simple
profiles or, most often, after excluding complex features (bumps and dips) at 
intermediate radii. The simultaneous two-component fit is based on a 
non-linear least-squares minimization algorithm to a function of two 
generalized exponentials. That is, we interactively fitted a function with 
{\em six} parameters, using as initial estimates the results from our 
r$^{1/4}$+exponential or exponential+exponential fits, above. The form that we
have adopted for the generalized exponential is the Sersic profile 
(\cite{sersic68})
\[ 
\Sigma(r)=\Sigma_0 e^{-(\frac{r}{h})^n}, \, n>0 
\]
$\Sigma(r)$ being the surface brightness at radius r, $\Sigma_{0}$ the central
surface brightness and {\em h} the scale length. The same formula can be 
written in terms of surface magnitudes
\[ 
\mu(r)=\mu_0 + 1.0857(\frac{r}{h})^n 
\]
(Note that in some studies the power \( \frac{1}{n} \) is used instead of n).
Smaller values of n lead to more cuspy central light distributions and 
shallower profiles outwards, while progressively larger values of {\em n} will
produce flatter central light distributions and truncated outer profiles 
(see Figure~\ref{f1}).
The generalized exponential function includes both the simple exponential case
({\em n}=1) and the De Vaucouleurs profile using the transformations:
\( \Sigma_0=2138\Sigma_e, \, \mu_0=\mu_e-8.325, \, h=2.89\times10^{-4}r_e \)
(the subscript {\em e} referring to the De Vaucouleurs parameters).

In fact, as shown by \cite{caon93}, the Sersic formula can be expressed in 
terms of the radius encircling half of the total luminosity and the 
corresponding surface brightness. In order to do this, one can express the 
total luminosity and the surface brightness introducing two coefficients 
that are functions of the exponent $n$. Then, for the range of values $n$
that we found, we can compute these coefficients by numerical integration
and find two approximate formulas for their dependence on $n$.
We have done this, combining Caon \etal's and our formulation, which gave
to a very good approximation the following transformations between $\mu,h$ 
and $\mu_{e},r_{e}$ for our data:
\[
\mu_e=\mu_0+2.5[0.868\frac{1}{n}-0.142]
\]
\[r_e=h [\frac{2.5(0.868\frac{1}{n}-0.142)}{1.0857}]^{1/n}
\]
We have used the errors in $\mu$ (calculated as described in \cite{thesis}) 
for (Gaussian) weighting of the data points, estimating the goodness of fit 
and calculating errors for the fitted coefficients.
Except for cases where the light profiles were relatively simple and the 
initial parameters well defined, we started by fixing the parameters for the 
best defined component and allowing the parameters for the second component to
vary until a good fit was achieved. Next we kept this set of parameters fixed 
and varied the other and iterated this process until a good solution was 
approached. At this point we allowed  all six parameters to vary freely and 
finally calculated the errors for the best fit values. For most objects the 
fitted range for the exponent {\em n} was 0-2, which is the range usually 
found in previous studies (see discussion in the next section). There were 
however a few cases of very complex profiles, for which one or both exponents 
{\em n} had to be kept fixed, in order for the fits to converge.

The resulting parameters, that is, the exponent {\em n}, the characteristic 
scale lengths {\em h} and the corresponding surface brightness levels $\mu$ 
are tabulated in Table~\ref{tab1}, where the subscripts {\em in} and {\em out}
denote, respectively, the inner (spheroidal) and outer (disk) components. 
Along with the fitted parameters, we also list in Table~\ref{tab2} the mean
ellipticity and position angle estimated at a certain isophotal radius (listed
in the same table). Since the outer isophotes for many of our objects are 
distorted and/or show tidal features (tails or one-sided arms) and strong 
spiral arms, it is not possible to define a common characteristic brightness 
level or radius for estimating the disk ellipticities and position angles. 
Instead, we inspected visually the direct images of all our objects and 
compared them to the fitted (elliptical) models, in order  to find the 
unaffected isophotes which are mostly located at the edge of the inner disk
(avoiding bars and inner rings). Inevitably, the parameters estimated this way
are somewhat subjective and should not be used to accurately estimate 
inclinations for instance, but they {\em are} indicative of the various 
subsamples and have a statistical usefulness.

The ellipse fitting procedure, outlined earlier, was applied to all of our 
objects for which either photometric information was available or which 
possessed well-resolved morphologies. In the Appendix we present surface 
brightness and colour profiles for all of them. In a number of cases it was 
not possible to achieve an accurate decomposition, because of the 
great complexity of the light distributions (\eg most of the mergers). Also 
we omitted profile decomposition for two objects with photometric information 
in the Warm sample, because of their small projected sizes. The total numbers 
of objects with analyzed light profiles, whose parameters are listed in 
Table~\ref{tab1}, are: 17 Seyfert 1s, 23 Seyfert 2s and 14 IR-Cold objects.
The parameters obtained from the profile decomposition of objects which have 
no photometric data, are valid for the scale lengths {\em h} and exponents 
{\em n}, but the surface brightness scales were chosen arbitrarily (for the 
decomposition purpose) and thus are omitted from Table~\ref{tab1} and from all 
subsequent plots. The information in this table is grouped together for 
each subsample (Warm Seyfert 1s, Warm Seyfert 2s and Cold galaxies) and, 
within each subsample, is ranging from early to late type host morphologies. 
The morphological classification, listed in Table~\ref{tab1}, is taken from the
literature (when available) and was revised and completed by us, using the 
revised Hubble classification system and the morphological index T 
(\cite{vaucoul91}).

\begin{table}
\dummytable\label{tab1}
\end{table}

\begin{table}
\dummytable\label{tab2}
\end{table}

\subsection{\bf Light Concentration Indices}

A characteristic of galaxy morphologies which is of considerable astrophysical
interest, particularly in relation to nuclear activity, is the degree to which
light is concentrated towards the central region of the galaxy. There are 
several parameters which can measure the central concentration. These are 
based on both the relative importance of the spheroidal component ($B$) to the
disk component ($D$) and on the ratio of different scale lengths.

(i) Index based on profile decomposition: In the case of a generalized
exponential law, we compute the integrated luminosity over the entire galaxian
surface to be
\[ I_{tot}=\frac{2\pi(1-\epsilon)h^2}{n}\Gamma(\frac{2}{n})I_0 \]
where \( \Gamma(m)=\int\limits_{x=0}^{\infty}e^{-x}x^{m-1}\,dx \) and 
\( \epsilon=1-\frac{b}{a} \) is the ellipticity. 
The relative importance of the two components is thus given by the ratio
\[ C_{I/O}=\frac{I_{tot(in)}}{I_{tot(out)}}=\frac{n_{in}}{n_{out}} \frac{\Gamma(\frac{2}{n_{in}})}{\Gamma(\frac{2}{n_{out}})} (\frac{h_{in}}{h_{out}})^2 10^{-0.4(\mu_{in}-\mu_{out})} \]
the subscripts {\em in} and {\em out} referring to the (inner) spheroidal and 
(outer) disk components, respectively. We parametrized the light concentration
using the following indices:

(ii) Index based on aperture photometry: We can define a similar index to 
$C_{I/O}$, based on aperture photometry results, this being the ratio of 
nuclear-to-disk ({\em N/D}) or nuclear-to-total ({\em N/T}) luminosities (as 
defined in Paper II).  In Table~\ref{tab3} we list $C_{I/O}$, {\em N/D} and  
{\em N/T}. 

(iii) Indices based on length ratios: These are based on some characteristic 
radii that are not affected by seeing effects in the inner regions or by sky 
subtraction uncertainties in the outer regions. Such indices can be defined to
characterize the light concentration at various scales and they should all be 
equivalent (\cite{okamura84}).

If \( L(r)=2\pi \int^{\infty}_0 I(r) \,r \,dr \) is the luminosity emitted 
within a radius {\em r} and \( k(r)=\frac{L(r)}{L_T} \) is the corresponding 
fraction of the total luminosity $L_{T}$, characteristic radii can be defined 
at:  $k(r_{1/5})$=0.20, $k(r_{1/4})$=0.25, $k(r_{1/2})$=0.5, $k(r_{3/4})$=0.75
and $k(r_{4/5})$=0.8. The most commonly used of these ratios is $r_{1/2}$, the
radius containing half of the {\em total} galaxian light (which is analogous 
to the De Vaucouleurs effective radius $r_{e}$ for the bulge component). 

\cite{vaucoul77} defined the concentration index 
\( c_{31}=\frac{r_{3/4}}{r_{1/4}} \).
This was found to correlate with morphological type, with $c_{31}$ decreasing 
from early to late type galaxies. Its theoretical value is 7.0 for a De 
Vaucouleurs profile and 2.8 for a pure exponential profile.
 
\cite{kent85} has defined the index 
\( c_{42}=\log{(\frac{r_{4/5}}{r_{1/5}})} \), which is equivalent to the 
Morgan concentration classification scheme (\cite{morgan58,morgan59}).
This index correlates with the \( \frac{B}{B+D} \) ratio 
(for \( \frac{B}{B+D} < 0.63\)) and with morphological type in the same sense 
as the $c_{31}$ index. Its theoretical values for a pure r$^{1/4}$ and an 
exponential profile, are similar to those of $c_{31}$. 

The surface brightness corresponding to the half-light radius $\mu_{1/2}$ 
shows a tight correlation with $c_{42}$ and $c_{31}$ (\cite{kent85}; 
\cite{vitores96b}), indicating that these
parameters describe equally well the galaxian light concentration. 
The usefulness of $\mu_{1/2}$ as a morphological type indicator is 
less well defined though, due to the large overlap between different types and
its generally large values in the case of emission line galaxies, 
independently of the host morphologies (\cite{vitores96b}). 
The large intrinsic dispersion in the correlations with Hubble type is a 
shortcoming in the use of all the above indices (\cite{vitores96a}).

The surface brightness corresponding to an outermost isophote such as 
$\mu_{24.5}$ shows no correlation with the above concentration indices or with
Hubble type. However, \cite{doi93} used it in combination with a new
concentration index $c_{in}(\alpha)$ to show that galaxies of different 
morphological types tend to segregate around different sets of values.
In particular, Seyfert type 1 and 2 galaxies are found to be well separated on
the $c_{in}$($\alpha$) vs $\mu_{24.5}$ diagram, Seyfert 1s being segregated in
the region $c_{in}$($\alpha$)=0.6-0.7, $\mu_{24.5}$=21.8-22.4 characteristic 
of early type galaxies, while Seyfert 2s lie in the region of later type 
galaxies with large scatter. 

We have computed many of the above light concentration indicators; the most
useful of them are tabulated in Table~\ref{tab3}. In order to calculate 
concentration indices an estimate of the {\em total} galaxian light
is needed. In \cite{thesis} we describe in detail how our ellipse fitting 
procedure works, including the calculation of total magnitudes integrating 
from the outermost fitted isophote to infinity. Here we use these
magnitudes to calculate the various characteristic radii and the
corresponding indices. In Table~\ref{tab3} we list the concentration 
parameters $C_{I/O}$, $N/D$, $N/T$, $c_{31}$, the characteristic radius 
$r_{1/2}$ and the diameter corresponding to the $\mu$=25 mag arcsec$^{-2}$ 
isophotal level (see Paper II).
Note that this table contains more objects than Table~\ref{tab1}, because it 
includes objects for which photometry is available (and thus concentration 
indices), but no profile decomposition was done.

\begin{table}
\dummytable\label{tab3}
\end{table}

\section{Profile Decomposition: Results and Discussion}

\placetable{tab4}

We shall now discuss the various structural parameters characterizing the 
galaxy bulges and disks for our three samples: their distributions, their 
correlations with each other and with additional observed galaxian properties.

\subsection{\bf Concentration Parameters vs Host and Seyfert Types}

We shall first consider the distributions of morphological index T (as defined
in RC3) and bulge-to-disk ratio, plotted in Figure~\ref{f2}. 

\subsubsection{\it Dependence on Morphological Type}

\placefigure{f2}

\placefigure{f3}

In general, a morphological classification is not precise and depends upon 
the type of data and the classification method used. The uncertainty in index 
T for two independent classifiers can range from 0.89 (RC3) to 2.2 
(\cite{lahav95}) with an average of 1.5 (\cite{jong96c}). Allowing for this 
uncertainty, which is about half a binsize in Figure~\ref{f2}, we 
find different trends in the distributions of T for the three subsamples: 

Warm Seyfert 1s tend to reside in earlier type hosts compared to Warm Seyfert 
2s, although both distributions peak at similar values. This tendency was 
noticed in a variety of previous studies, where Seyferts were also found to 
preferentially have spiral or barred spiral morphologies or other kinds of 
disturbances (\eg rings) 
(\eg \cite{adams77,wehinger77,simkin80,dahari84,kenty90}). 
In these respects, our IR-Warm Seyferts do not appear to be different than 
their optically selected counterparts. Our Cold sample galaxies show a clear 
shift towards later type hosts, although an accurate classification for these 
objects is difficult given that most are members of closely interacting 
systems, thus often severely distorted. 
Consequently, in what follows we denote with T$\ge$10 the systems with severe
distortions, recent mergers or strong tidal features (\eg one-sided arms) 
whose main body cannot be classified morphologically. 

\subsubsection{\it Dependence on Bulge to Disk Ratio}

\begin{figure}
\epsscale{1.}
\plotone{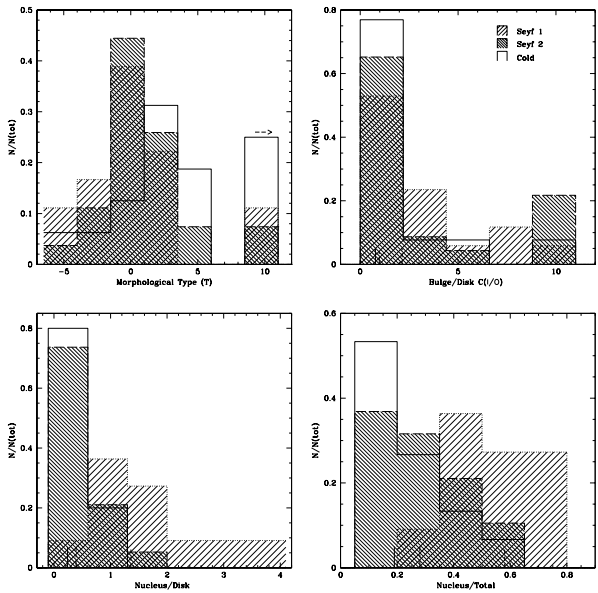}
\caption{Distributions of the morphological types T and the ratios Bulge/Disk (fitted radial profiles), Nuclear/Disk and Nuclear/Total (aperture photometry) for the Warm Seyfert 1 and 2 and the Cold galaxy samples. The vertical bars in the lower x-axes indicate the median values for each sample (also listed in Table~\ref{tab4}): short-dashed for Seyfert 1s, long-dashed for Seyfert 2s and solid for the Cold sample.\label{f2}}
\end{figure}

The index $C_{I/O}$ is, as defined in the previous section, equivalent to the 
commonly used bulge-to-disk ratio (\(\frac{B}{D}\)) and its distribution for 
the various samples is shown in Figure~\ref{f2}. The $C_{I/O}$ 
distributions of the Warm Seyfert 2 and Cold samples are similar, while the 
Seyfert 1s show a tail to higher values with a significantly larger median 
(Table~\ref{tab4}). The F-test and Student's t-test however show no 
statistically significant differences in the variances or means of the three 
samples. Let us compare our sample's $C_{I/O}$ with that for normal galaxies:
A classification scheme established from a sample of field galaxies from the 
HST Medium Deep Survey through a classical De Vaucouleurs+exponential 
decomposition (\cite{schmidtke97}), predicts 1.7\( \le \frac{B}{D} \le \)10
for bulge-dominated systems, 0.5\( \le \frac{B}{D} \le \)1.7 for intermediate 
types (S0s) and \( \frac{B}{D} \le \)0.5 for disk-dominated and pure disk 
galaxies. (Thus, $C_{I/O}$ larger than 10 in Figure~\ref{f2},
indicate almost pure-bulge systems). We caution however that the above values 
of \(\frac{B}{D}\) are uncertain and the error can be as large as 73\% of its 
value (\cite{schmidtke97}). Even though uncertainties in our $C_{I/O}$ ratio 
could be of similar magnitude, but the results depicted in 
Figure~\ref{f2} are statistically significant and in fact agree with
the normal galaxy classification: Seyfert 1s (earlier type hosts) are 
preferentially bulge-dominated systems while in Seyfert 2 (later type) hosts 
the disks are predominant. A similar result was reached by us earlier, in 
Paper II, from aperture photometry when comparing nuclear and disk or total 
magnitudes. 

We plot in the same Figure~\ref{f2} the distributions of these 
luminosity ratios, Nuclear/Disk and Nuclear/Total (from Paper II). There
is a clear difference between the distributions of Seyfert 1s and Seyfert 2s,
this being mainly due to the larger AGN contribution in the former. 
The K-S test attributes a statistical significance of 99\% to the hypothesis
that the two distributions are different. On the other hand, there is a
striking similarity in the distributions of the Warm Seyfert 2 and Cold 
galaxies (although at statistical level $<$95\%). The median
Nuclear/Total ratio (Table~\ref{tab4}) for Seyfert 1s is typical for S0s-S0/a 
galaxies, for Seyfert 2s characteristic of Sab-Sb types and for 
the Cold sample is characteristic of Sbc-Sc galaxies (\eg \cite{kent85}). 
While the integrated nuclear magnitudes used in these ratios are dominated by 
the bright AGN (at least for Seyfert  1s), the bulge component was fitted 
{\em excluding} the nuclear (2 kpc) region and thus the $C_{I/O}$ index should
be unaffected by the AGN. Consequently, the difference established above 
between the Seyfert type 1 and 2 host morphological types and light 
concentration, should not be biased by torus orientation/obscuration effects 
and thus be an {\em intrinsic} property of these galaxies. 

\begin{figure}
\epsscale{0.7}
\plotone{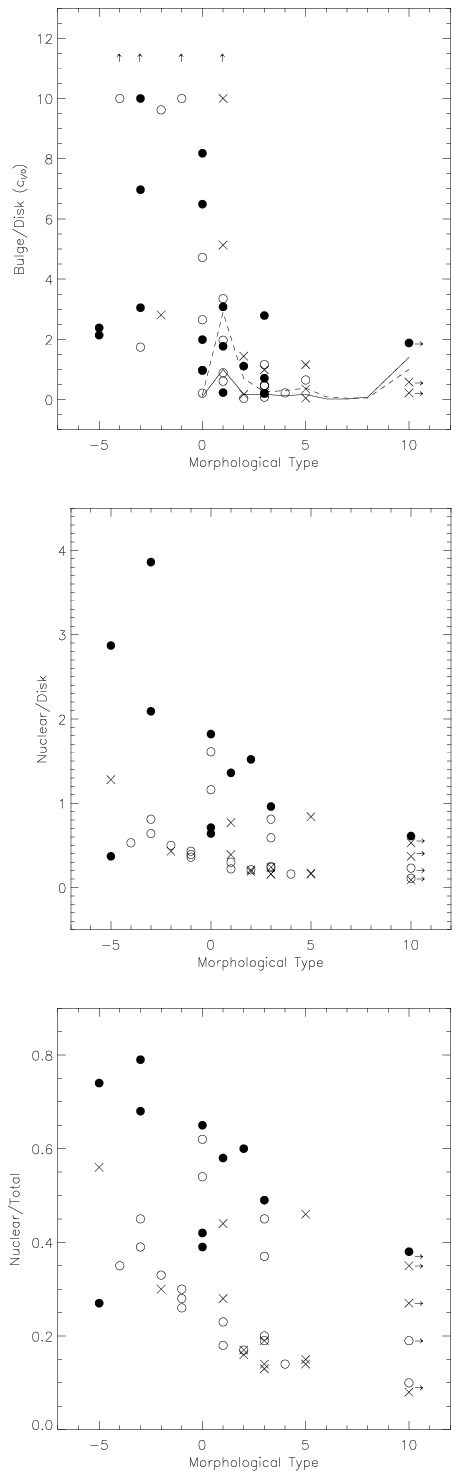}
\caption{The distribution of Bulge/Disk, Nuclear/Disk and Nuclear/Total light ratios as a function of morphological type T. Filled dots represent Seyfert 1s, open dots Seyfert 2s and crosses Cold galaxies. Horizontal arrows indicate T=10, arbitrarily assigned to objects with complex or amorphous structures (usually mergers or tidal) while the vertical arrows indicate lower limits for the Bulge/Disk ratios. Overplotted lines indicate median $B/D$ values for a sample of face-on spirals from De Jong 1996c (solid/dashed lines indicate $B$-band/$K$-band data, respectively).\label{f3}}
\end{figure}

A correlation between $B/D$ ratio and morphological sequence is suggested by 
previous studies, although morphological classification and bulge/disk 
decomposition uncertainties introduce large discrepancies 
(\eg \cite{kent85,simien86,andredakis94,jong96c}). If true, such a correlation
is important because it indicates a common formation process for the galaxy 
bulges and disks. 
In Figure~\ref{f3} we plot Bulge/Disk, Nuclear/Disk and Nuclear/Total
ratios against morphological type for all our objects. There is a trend, with 
large scatter, for these ratios to correlate with host type morphology 
(particularly in Seyfert 1s) in the expected sense, that is, larger ratios 
towards earlier type hosts. The scatter in $C_{I/O}$ is particularly large for
T$\le$0, due to the difficulty in distinguishing real S0s and ellipticals from
later-type objects, with low surface brightness disks. 
The Seyfert 2s and in particular the Cold galaxies show frequently disturbed 
morphologies and tidal features that lead to uncertain classifications and 
explain the scatter in the above diagrams. This is the case for some Seyfert 1
galaxies too, for instance IRAS 23016+2221 (the filled circle in 
Figure~\ref{f3} with systematically smaller ratios for its assigned 
morphological type).
In the same Figure~\ref{f3} we indicate the median value of $B/D$ 
ratio vs morphological type for the sample of face-on spirals of 
\cite{jong96c} (the solid/dashed lines indicate $B$-band/$K$-band data, 
respectively). Our ratios are computed at an intermediate wavelength (R band)
and are consistent on average with these lines.

\placefigure{f4}

\subsubsection{\it Dependence on Concentration Parameters}

\begin{figure}
\epsscale{1.}
\plotfiddle{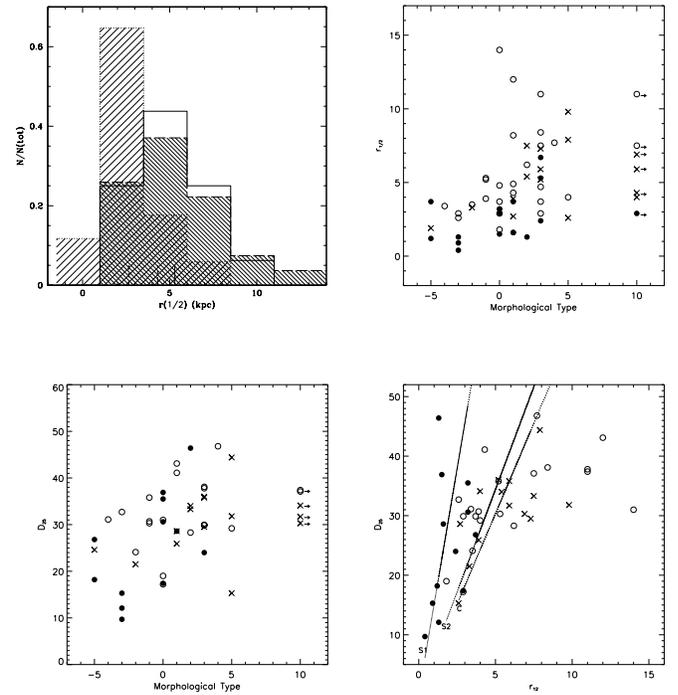}{400pt}{0}{47}{47}{-150}{-70}
\caption{Distributions of half-light radii $r_{1/2}$  for the different samples and as function of morphological type (upper panels). Host diameters at $\mu_{B}$=25 mag arcsec$^{-2}$ as a function of morphological type and the correlation between the two scale lengths (lower panels). Overplotted on the lower right panel are lines of constant ratio between the two scale lengths (the observed median ratio for each sample). Other symbols are the same as in Figures 2 and 3.\label{f5}}
\end{figure}

\placefigure{f5}

\begin{figure}
\epsscale{1.}
\plotfiddle{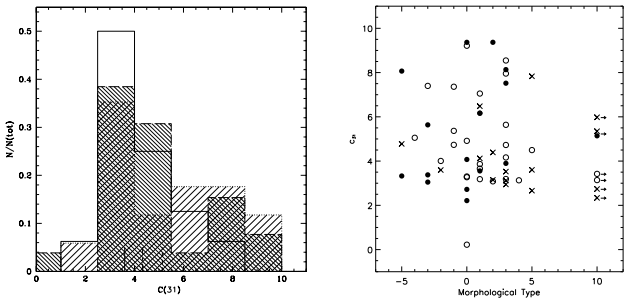}{400pt}{0}{55}{55}{-120}{0}
\caption{Distributions of the concentration index $c_{31}$ for the different samples and as function of morphological type. Symbols are the same as in Figures 2 and 3.\label{f4}}
\end{figure}

The concentration parameter $c_{31}$ is shown in Figure~\ref{f4}. 
This index is more likely (compared to $C_{I/O}$) to be affected by the 
presence of bars, inner rings, or other structures in the ``intermediate'' 
zone of light profiles. These features seem to be quite common in our 
sample galaxies and are likely to introduce large scatter in any existent
correlations. This is indeed what we see in Figure~\ref{f4}. It is 
interesting that \cite{vitores96b} have similarly found a large range of 
$c_{31}$ (2.65, 7.21) for their Seyfert sample.  We thus conclude that 
$c_{31}$ is not very useful for objects with disturbed morphologies.The median
values, listed in Table~\ref{tab4}, are characteristic of S0/a types or earlier 
for the Seyfert 1 sample, Sa-Sb types for the Seyfert 2 sample and Sbc types 
or later for the Cold sample. 

$r_{1/2}$, the radius within which half of the {\em total} galaxian light is 
emitted, shows better discriminating power. In Figure~\ref{f5} we see
a clear difference in the distribution of this parameter between the various 
samples: Seyfert 1s have smaller $r_{1/2}$ compared to Seyfert
2s and Cold galaxies.  This is certainly due to the nucleus dominating the 
total light of Seyfert 1s.  The F-test and Student's t-test show that the 
Seyfert 1 and 2 samples have different variances and means at significance 
levels 0.008 and 6.3E-5, respectively. The K-S test lends 
statistical significance of $\sim$99.8\% that the Seyfert 1 $r_{1/2}$
distribution is different from any of the other two (sub)samples.
On the other hand, the Seyfert 1 and Cold galaxy $r_{1/2}$ distributions match
at significance level 96\% and also have statistically similar variances and 
means.

We also find a correlation between $r_{1/2}$ and host type T, which is 
statistically significant only for the Seyfert 2 sample (significance 0.009, 
with the Spearman's $\rho$ or Kendalls's $\tau$ rank correlation). 
However $r_{1/2}$ is also affected by the presence of bars or other central 
structures, which is the case of the three deviant Seyfert 2s towards larger 
$r_{1/2}$ values (for their assigned morphological type) in 
Figure~\ref{f5}. 
In the same figure we plot the diameters of the $\mu_{B}$=25 mag
arcsec$^{-2}$ isophote, $D_{25}$, as a function of T. There is an obvious 
trend for the two quantities to correlate, but with large scatter. The
Spearman's rank correlation test shows no correlation at a statistically 
significant level, for any of the three samples.
 
The trends/correlations found above between size parameters and morphological 
type T, suggest that later type hosts tend to have larger scale-lengths 
{\em and} sizes. 
Is this in the sense of the expected linear correlation between $r_{1/2}$ and 
$D_{25}$, or does it indicate that late-type hosts tend to be more diffuse 
(less centrally concentrated)? The $r_{12}$ vs $D_{25}$ plot in Figure~\ref{f5} 
shows a correlation between the two quantities (with large scatter) for the 
three samples. Overplotted are lines of constant ratios between the two scale 
lengths (the observed median ratio for each sample), roughly indicating the 
path that points would follow if all galaxies in each sample had similar 
light profile shapes and $\mu_{0}$. Most Seyfert 1s follow such a correlation and this most 
likely indicates the nuclear dominance of their total light. For Seyfert 2 and
Cold galaxies the slope of the relation becomes flatter at large radii/sizes 
(and thus, as we have shown, later host types). We conclude that the Warm 
Seyfert 2 and the Cold galaxies have indeed shallower light profiles compared 
to the Warm Seyfert 1s.

Previous studies have shown that Seyferts are larger than normal spirals or 
other emission line galaxies. Measured at the 24 mag arcsec$^{-2}$ isophotal 
level, mean Seyfert diameters are found to be in the range 22.5 kpc 
(\cite{salzer89}), 25$\pm$10 kpc (\cite{kenty90}), or 36$\pm$15 kpc 
(\cite{vitores96b}). The mean diameters (measured at the 25 mag arcsec$^{-2}$ 
isophotal level) of our Warm Seyferts are in the same range (see Paper II):
27 (median 25.4) kpc for Seyfert 1s and 33.3 (median 31) kpc for Seyfert 2s. 
This is also the case for Cold sample objects, mean $D_{25}$=30.9 (median 
31.7) kpc, although most of them do not harbour AGN. We conclude that 
IR-active galaxies tend to be brighter and larger at a given isophote than 
normal spirals, independently of their nuclear activity stage.

\subsubsection{\it Dependence on Ellipticities}

\begin{figure}
\epsscale{1.}
\plotfiddle{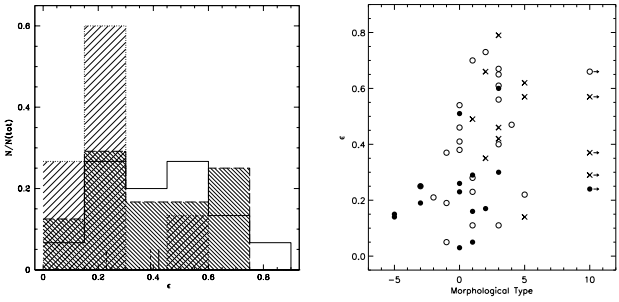}{400pt}{0}{55}{55}{-120}{0}
\caption{The distribution of ellipticities, $\epsilon$, for the three samples and $\epsilon$ as a function of morphological type. Symbols are the same as in Figures 2 and 3.\label{f6}}
\end{figure}

\placefigure{f6}

In Figure~\ref{f6} we show the distribution of ellipticities
$\epsilon$ for our (sub)samples and also plot the ellipticity as a 
function of morphological type. The distributions are similarly flat for 
Seyfert 2s and Cold galaxies, ranging from $\sim$0-0.8, while the range of
ellipticities is much narrower for 
Seyfert 1s, peaking at $\epsilon\approx$0.2 (see also Table~\ref{tab4}). 
Previous findings that Seyferts have in general ellipticities $<$0.5 
(\cite{keel80,kenty90,vitores96b}) is confirmed only for Seyfert 1s and 
could be due to these being predominantly face-on systems. However, if the 
ellipticities presented here are indicative of the galaxy's inclination, their
flat distribution would also imply that Seyfert 2s are {\em not} predominantly
edge-on systems but are oriented randomly on the sky. Another possible 
explanation for the difference in apparent $\epsilon$ between our samples is 
that, if galaxies are thin disks then the ellipticity distribution is expected
to be flat, from inclination effects alone. On the other hand, galaxies with 
strong bulge components would have a distribution of ellipticities peaking 
around zero. (In this case one would expect to find a well-defined correlation
between $C_{I/O}$ and $\epsilon$, which however is not the case.

As we discussed in the previous section, the need to avoid distorted 
isophotes, that are
present predominantly in the Seyfert 2 and Cold samples, introduced non 
uniformity in the measures of $\epsilon$. It is thus likely that this 
procedure has introduced the large scatter in $\epsilon$ for these two 
samples.

\subsection{\bf Bulges and Disks}

\subsubsection{Distributions of Characteristic Parameters}

\begin{figure}
\epsscale{1.}
\plotone{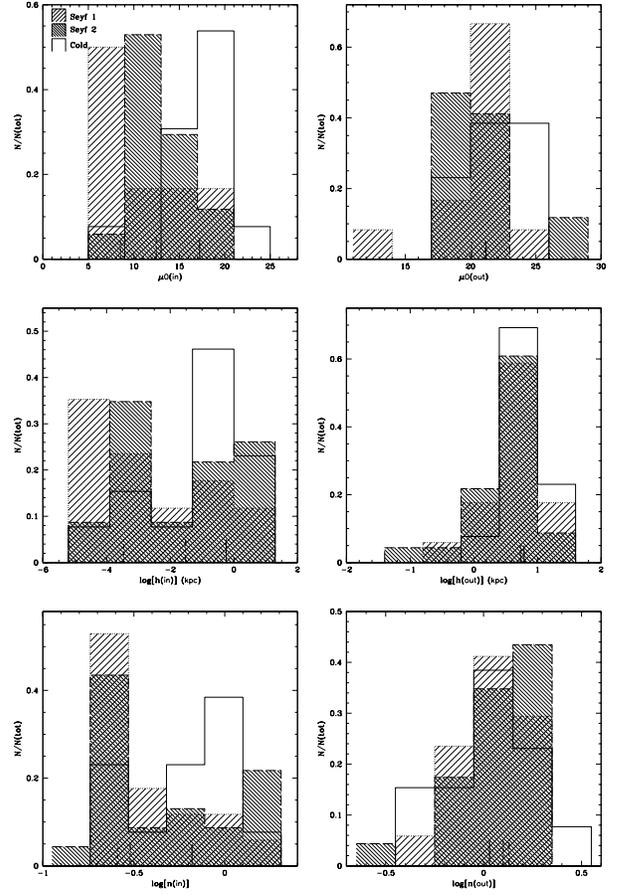}
\caption{The distributions of the fitted parameters of two generalized exponentials, for the bulge(inner) and disk(outer) component, of the radial light profiles. The vertical bars in the lower x-axes indicate the median values for each sample (also listed in Table~\ref{tab4}): short-dashed for Seyfert 1s, long-dashed for Seyfert 2s and solid for the Cold sample.\label{f7}}
\end{figure}

In Figure~\ref{f7} we show the distributions of the bulge (inner) and
disk (outer) best fitting parameters, calculated as described in the previous 
section. We will show that the bulge parameters differ significantly between
the three (sub)samples whereas there are less marked differences in their disk
properties. More precisely, we find that: 

\placefigure{f6}

(i) Seyfert 1 bulges have systematically larger central surface brightnesses 
than Seyfert 2s, although the two subsamples span similar ranges, roughly 
$\mu_{in}$=5-20 mag arcsec$^{-2}$. The median $\mu_{in}$ values are
given in Table~\ref{tab4} and correspond to $\mu_{e}$=17.64 
and 18.77 mag arcsec$^{-2}$, respectively. The Cold galaxies have fainter 
bulges, with a median $\mu_{in}$ equivalent to $\mu_{e}$=19.85. The F-test 
shows not significantly different variances for the three (sub)samples, but 
the Student's t-test shows that Cold galaxies have significantly different 
mean compared to the Seyfert 1 and 2 galaxies (significance 0.001 and 0.01, 
respectively). The K-S test shows none of the three (sub)samples to have 
similar distributions.

(ii) The bulge scale lengths span a large range and have similar, flat 
distributions for the two Seyfert subsamples, but a smaller median for 
Seyfert 1s (see Table~\ref{tab4}). However, the Seyfert 1 bulges have in 
general steeper profiles (smaller $n_{in}$), thus the median half light radii 
(approximated by the relation given in Section 2.2) are similar for the
two Seyfert subsamples: $r_{e}$=1.35 and 1.06 kpc for type 1s and 2s, 
respectively). There is a bimodal character in the $h_{in}$ distributions
for all samples, around values that correspond to {\em n}=0.25 and 1, that is,
to De Vaucouleurs and simple exponential profiles. Since the simultaneous two 
component fit was done using initial parameters from two independent
bulge and disk fits assuming either an $r^{1/4}$ or $r^{1}$ law (see
previous section), there might be a best-fit ``bias'' towards these values in 
the final results.
The Cold sample bulge scale-lengths are shifted to larger values with respect 
to the Warm samples (equivalent median $r_{e}$=2.14 kpc) and the variance of 
its distribution of $h_{in}$ values is significantly different than the 
Seyfert 1 and 2 subsamples (significance of F-test 3.2E-7 and 1.4E-5,
respectively). 

(iii) There is a marked difference in the distributions of exponents $n_{in}$ 
between the Warm and Cold samples. Although this parameter spans a remarkably 
large range for all three samples, Seyfert 1 and 2 galaxies have similar bulge
profiles with a median value of {\em n}=0.26-0.30, that is very close to a De 
Vaucouleurs $r^{1/4}$ law. The tendency seen in Figure~\ref{f7} for the 
Seyfert 1 bulges to be steeper is not confirmed at a statistically significant
level. The Cold galaxy bulges are flatter (larger $n$), with a median 
{\em n}=0.66 and a bimodal distribution (around $n_{in}$=0.25 and 1). 

(iv) The distributions of the half-light radius $r_{e}$ and corresponding 
surface brightness $\mu_{e}$, show systematic shifts between the three 
(sub)samples, although less pronounced compared to the $h_{in}$ and $\mu_{in}$
distributions. There is no bimodality in the $r_{e}$ distributions for any of 
the samples. $\mu_{e}$ is 
brighter for Seyfert 1s compared to Seyfert 2s and the latter brighter than
Cold galaxies. $r_{e}$ spreads between $\sim$0.1-10 kpc for the three samples 
( which are typical values for ellipticals and bulges of disk galaxies), with 
a peak at $\sim$ 1kpc for the Warm sample. The Cold sample $r_{e}$ 
distribution is flatter and shifted to larger values compared to the Warm 
sample. 

It is instructive to compare our results for the bulge parameters with 
previous studies attempting to parametrize spiral galaxy bulges:
Profile decompositions using a typical De Vaucouleurs+exponential combination 
yield median values of $\mu_{e}$=20.8 mag arcsec$^{-2}$ for S0s and 22.5 mag
arcsec$^{-2}$ for Sc types (\cite{kent85}). A sample of emission line galaxies
(\cite{vitores96b}) shows mean $\mu_{e}$=22.5$\pm$1.6 mag arcsec$^{-2}$, 
$r_{e}$=2.1 kpc and \( \frac{B}{D} \)=0.75. Thus, the Warm Seyfert galaxies 
have larger characteristic bulge brightness for similar host morphological 
types. \cite{andredakis95} and \cite{jong96b} used 2D
fitting techniques for their samples of spiral galaxies and applied 
a generalized exponential to the bulge component in combination with a simple 
exponential for the disk. They found a large variety of bulge shapes, the 
parameter {\em n} (as defined by us in Section 2) ranging between $\sim$0.2-2.
\cite{jerjen97} applied generalized exponentials to ellipticals and found an 
even larger range for $n\approx$0.08-2.5. Thus the Warm Seyfert 
bulges have on average similar profile shapes as normal galaxies for similar
morphological types.
 
(v) In Figure~\ref{f7} the disk central surface brightnesses show 
quite similar distributions for the three (sub)samples (mean range of
$\mu_{out}\sim$18-24 mag arcsec$^{-2}$), with similar variances and means.
Using the mean disk $(B-R)$ colours for each sample from Paper II, we find a 
range of $\mu_{Bout}$=19.1-25.1 mag arcsec$^{-2}$ for the three samples, with
median values 22.01, 21.24, 22.35 mag arcsec$^{-2}$ for the Warm Seyfert 1, 
Seyfert 2 and Cold galaxies, respectively. Given the relatively large errors 
associated with the fitted parameter $\mu_{out}$ (see Table~\ref{tab4}), these 
median values are in agreement with the remarkably uniform, {\em inclination 
corrected} central disk brightness first noted by Freeman: 
\( B_0(c)=21.65\pm0.30(\sigma)B\mu \) (\cite{freeman70,boroson81,kent85}).
However, the relatively large range of $\mu_{B}$ that we found for our samples
argues against a uniform disk central surface brightness (even allowing for 
the large errors associated with $\mu$).

The significance of Freeman's findings about exponential disks has been 
questioned by various studies since then. \cite{kormendy77b} showed that the 
``universal'' value of brightness is probably due to selection effects
against faint disks and to underestimation of the bulge contribution at outer 
radii. This conclusion was supported also by later studies 
(\cite{davies90,phillips93}). Indeed, Kormendy fitted the brightness 
distributions of bulges with a modified Hubble law, and found 
\( B_0(c)=21.70\pm0.23(\sigma)B\mu \) for compact galaxies and 
\( B_0(c)=22.78B\mu \)-22.68B$\mu$ for normal galaxies, which is closer to the
mean value of our Cold sample. Other explanations have also been put forward 
for the apparent uniqueness of $\mu_{0}$: (a) Selection effects which 
discriminate against very compact galaxies and galaxies with low surface 
brightness (\cite{vaucoul74,disney76,allen79}). Indeed, \cite{romanishin83} 
found a mean $\mu_{0}$=22.74 $B$ mag arcsec$^{-2}$ for low surface brightness 
galaxies and \cite{kruit87} obtained $\mu_{0}$=22.5 $B$ mag arcsec$^{-2}$ for 
his sample of spiral galaxies. (b) Dust extinction effects in the $B$ passband
(\cite{jura80,valentijn90,peletier94}). (c) The $B$ passband that Freeman used
to establish his relation is dominated by young stars that represent only a 
few percent of the total stellar mass in a galaxy (\cite{jong96b}). De Jong 
finds that there is no preferred value for the disk central surface brightness
of spirals, but only an upper limit and shows that none of the above 
alternatives {\em alone} is enough to explain Freeman's law. \cite{yee83} 
found that Seyferts have similar colours and other disk parameters as normal 
galaxies, but with a tendency for higher central surface brightness (21.3 $B$
mag arcsec$^{-2}$) compared to normal spirals. His results agree with ours for
the Warm Seyfert 2 sample. \cite{kenty90} found a fainter mean value 
(21.9 $B$ mag arcsec$^{-2}$) for his sample of Seyfert galaxies, which is
consistent with the average of our Warm Seyfert sample.

(vi) We find similar disk scale length distributions for the three samples, 
with a mean range $h_{out}$=1-16 kpc. The Student's t-test shows no 
significant difference between the three samples means, but the Seyfert 2 and 
Cold samples have different variances (significance 
0.05 of the F-test). \cite{freeman70} found the disk scale lengths to vary 
between 2 and 10 kpc in S0-Sbc galaxies and between 2 and 5 kpc in Sc-Im 
galaxies. Similarly, \cite{jong96b} found that face-on spirals tend to have 
disk scale lengths in the range 1-10 kpc. \cite{kormendy77b} found mean scale 
lengths 6.5-8.5 kpc for his late type spirals, while \cite{kruit87} gives a
mean value of only $\sim$1.5 kpc for his sample of spirals. For a sample of 
(mostly late type) emission line galaxies, \cite{vitores96b} measured 
$h$=3.2$\pm$2.8 kpc and \cite{kotilainen94} also found small mean scale 
lengths $\sim$2.5 kpc for their Seyfert sample. We conclude that our 
IR-selected galaxies have similar median disk scale lengths as normal spirals, 
independently of their nuclear activity type (that is, both Warm and Cold 
samples).

(vii) The disks of our sample galaxies were fitted with a large variety of 
profiles, $n_{out}\approx$0.6-1.9 with a median value of 1.3 for the Warm 
Seyferts and of $\sim$1 for the Cold galaxies. Differences between their means
or variances are not statistically significant.

\subsubsection{Correlations with T and Light Concentration Indices}

The question arises whether any (or all) of these parameters are related to
the host galaxy morphological type. Let us first briefly summarize the 
literature on the subject. \cite{kent85} applied a classical $r^{1/4}$ bulge +
exponential disk decomposition to a galaxy sample covering a large range of 
morphological types and found that the bulge surface brightness decreases 
towards later Hubble types (i.e. with decreasing total bulge luminosity). He 
found (as expected) decreasing \( \frac{B}{D} \) or \( \frac{B}{T} \) 
ratios as a function of morphological type, which he showed to be due to 
decreasing $\mu_{e}$ rather than decreasing bulge size (however, 
\cite{vitores96b} finds no evidence for this). \cite{jong96c} noticed a 
similar tight correlation between host morphological type and bulge central 
surface brightness but not with bulge scale length. \cite{andredakis96} and 
\cite{jong96c} applied generalized exponentials to the bulges of spiral 
galaxies and found a good correlation between {\em n} and morphological
 type (or \( \frac{B}{D} \) ratio) in the sense that, S0 bulges resemble an 
$r^{1/4}$ law, Sa-Sb bulges are best fit with {\em n}=0.5, while bulges of 
late type spirals are closer to a simple exponential {\em n}=1 with possible
values as high as $n\sim$2. Similar correlations for the disk parameters are 
uncertain: a tendency was found for the central surface brightness to decrease
with morphological type, but with large scatter, whereas there is no obvious 
correlation between disk scale length and morphological type. The apparent 
lack of correlation between bulge or disk scale lengths and Hubble type, if 
true, would mean that the Hubble type is scale-free (\cite{jong96c}).

\begin{figure}
\epsscale{1.}
\plotone{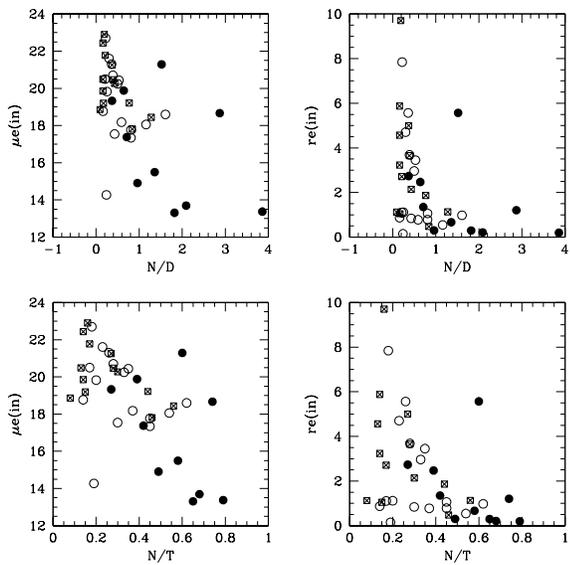}
\caption{Bulge parameters (outside the central 2 kpc) plotted against light concentration indices. Filled/open circles indicate Warm Seyfert types 1 and 2, respectively and crossed triangles represent the Cold sample galaxies.\label{f8}} 
\end{figure}

\placefigure{f8}

We searched for similar correlations in our data but found {\em no}
significant correlation
between the bulge or disk parameters and morphological type for any of the
samples. Given the uncertainties associated with T (discussed above) this 
result is not surprising. Thus, we searched for correlations with other
light concentration indices. We find a tendency for larger $\mu_{e}$ with 
light concentration, as expressed by $C_{I/O}$, for Seyfert 1s. There are also
correlations between $\mu_{e}, r_{e}$ and $N/T$, $N/D$, shown in
Figure~\ref{f8}. The scatter is large though and the correlations are 
better defined for the Cold sample. In addition, the Cold sample 
$\mu_{e}, r_{e}$ parameters correlate with host size $D_{25}$, in the sense
of fainter surface brightnesses and larger half light radii for larger hosts.
$\mu_{e}$ (calculated outside the central 2 kpc) also correlates with 
nuclear (within 2 kpc) luminosity for the Cold galaxies. The correlations 
found above for the Cold sample indicate a smooth distribution of light
form the disk to the center, in the absence of an active nucleus, in these
objects. The absence of any significant correlation between bulge and nuclear 
parameters for the Warm Seyfert sample, on the other hand, is a further
confirmation that the bulge parameters are not significantly affected by the 
AGN. We searched for similar correlations for the disk parameters of our
objects. We only find a tendency for Seyfert 1 disks to have larger $n_{out}$ 
(that is, shallower inwards - truncated outwards profiles) in 
nucleus-dominated objects (larger N/D and N/T ratios) and a tendency for 
Seyfert 2 bulges to have $n_{in}>$0.25 only in disk-dominated objects (smaller
N/D and N/T ratios). Both these relations are in the expected sense, also
found in normal (non-AGN) galaxies. 

On the plots of 
Figure~\ref{f8} we see clearly the different loci in parameter space 
that the three (sub)samples occupy: Cold galaxies have fainter $\mu_{e}$, 
larger $r_{e}$ and smaller light concentration indices, the opposite being 
true for the Warm Seyfert 1 sample. The Warm Seyfert 2 galaxies occupy 
intermediate loci in these plots, tentatively suggesting a transition between 
the above two samples.
The results outlined so far, indicate significant differences in bulge
parameters between the two Seyfert types, which can not be accounted for 
by the orientation/obscuration unification model (at least in its simplest 
form, involving the torus orientation).

\subsubsection{Inter-Correlations of Bulge and Disk Parameters}

\begin{figure}
\epsscale{1.}
\plotfiddle{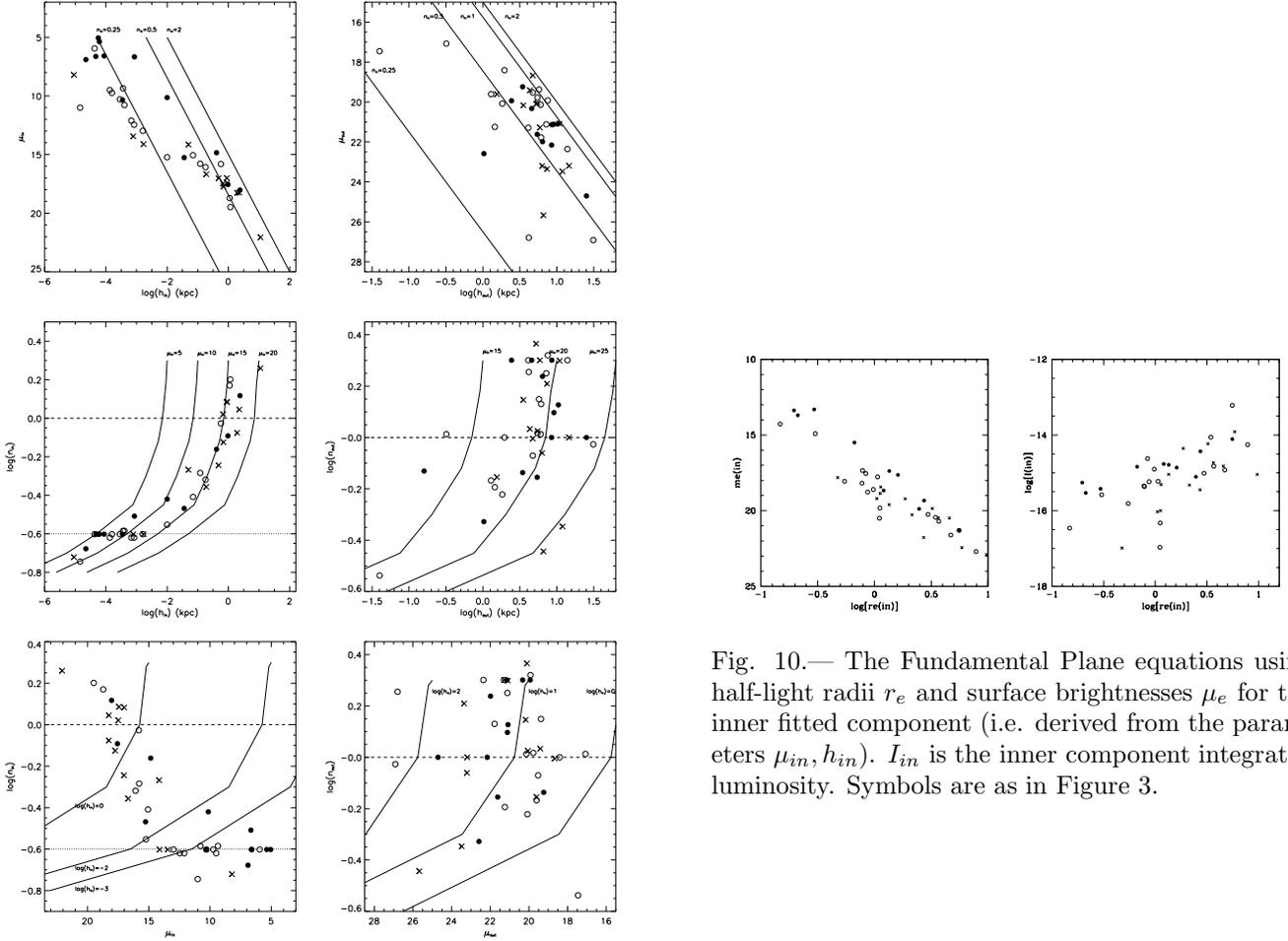}{400pt}{0}{85}{85}{-125}{20}
\caption{Fitted bulge (left panels) and disk (right panels) parameters, plotted against each other. Symbols are as in Figure 3. The full lines indicate constant (bulge or disk) luminosity,the dotted line a De Vaucouleurs ({\em n}=0.25) and the dashed line a simple exponential law ({\em n}=1).\label{f9}}
\end{figure}

\placefigure{f9}

In Figure~\ref{f9} the fitted parameters are plotted against each other,
for each of the bulge and disk components. We find some interesting 
correlations:

(i) The bulge scale length and central surface brightness (upper left panel) 
show a good correlation with each other, for all three samples. We used the 
median total bulge luminosity to overplot (solid) lines 
of constant luminosity for a range of $n_{in}$ values. For larger total 
luminosities the lines move parallel and to the right. Two main conclusions 
can be drawn from this plot: (a) There is a limited range in {\em total} bulge
luminosities. The brightest objects follow the line of constant luminosity for
$n_{in}$=0.25 and then move towards larger $n_{in}$ values. This is equivalent
to the bimodality in the $h_{in}$ and $n_{in}$ distributions that we found 
earlier. (b) There seems to be an upper and a lower limit in the ($\mu_{in}$, 
$h_{in}$) space: there are no objects with large bulge scale-lengths and high 
central surface brightnesses or, on the opposite, short scale lengths and low 
central surface brightnesses, although the latter could be partially due to a 
selection effect against fainter galaxies. 
The similar plot for disk parameters (upper right panel) shows some 
correlation, most of the points lying between the lines of equal luminosity, 
corresponding to $n_{out}$=0.5-1. One can also define upper and lower limits 
here, as in the case of bulge parameters.
 
\begin{figure}
\epsscale{3}
\plottwo{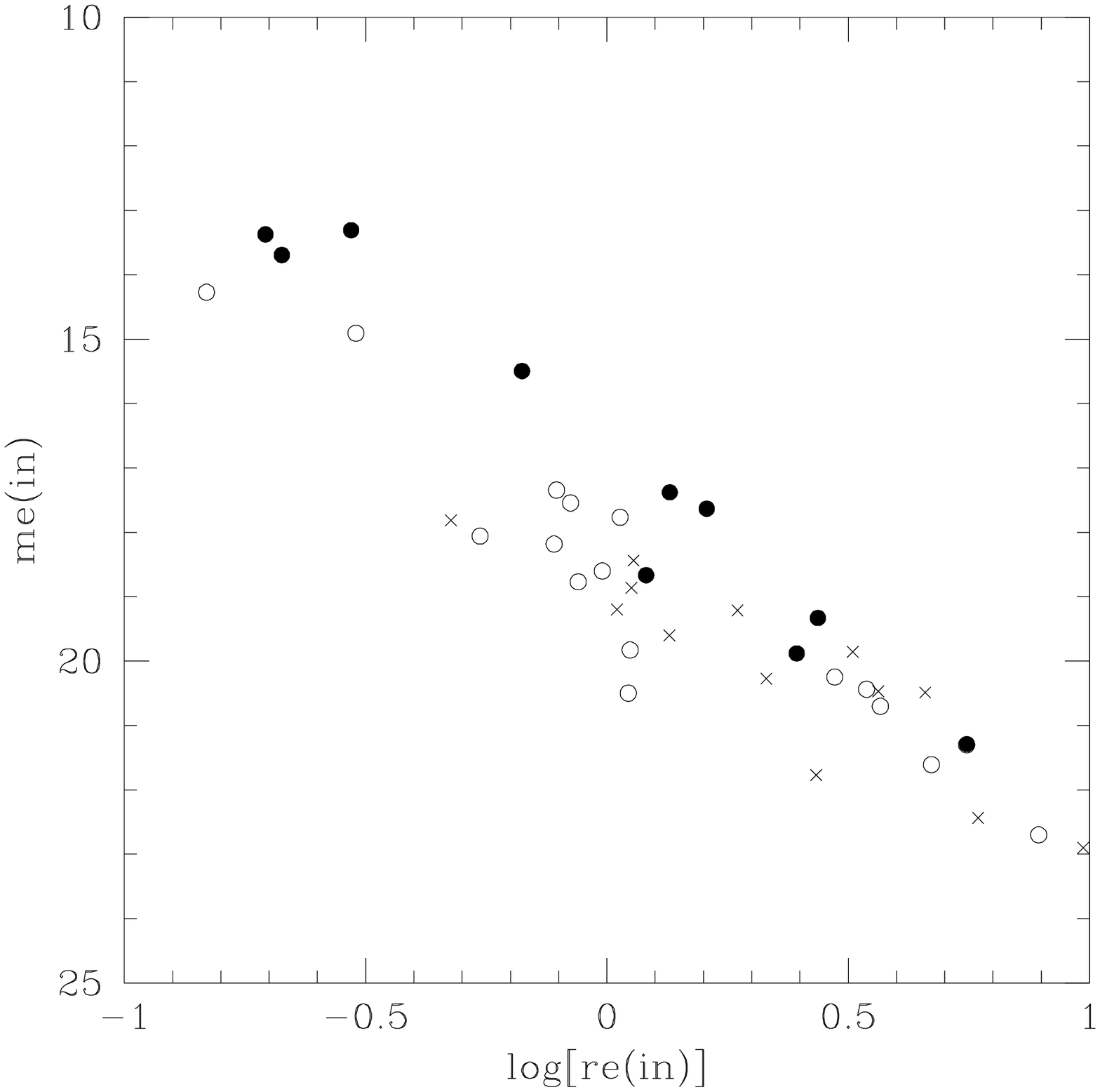}{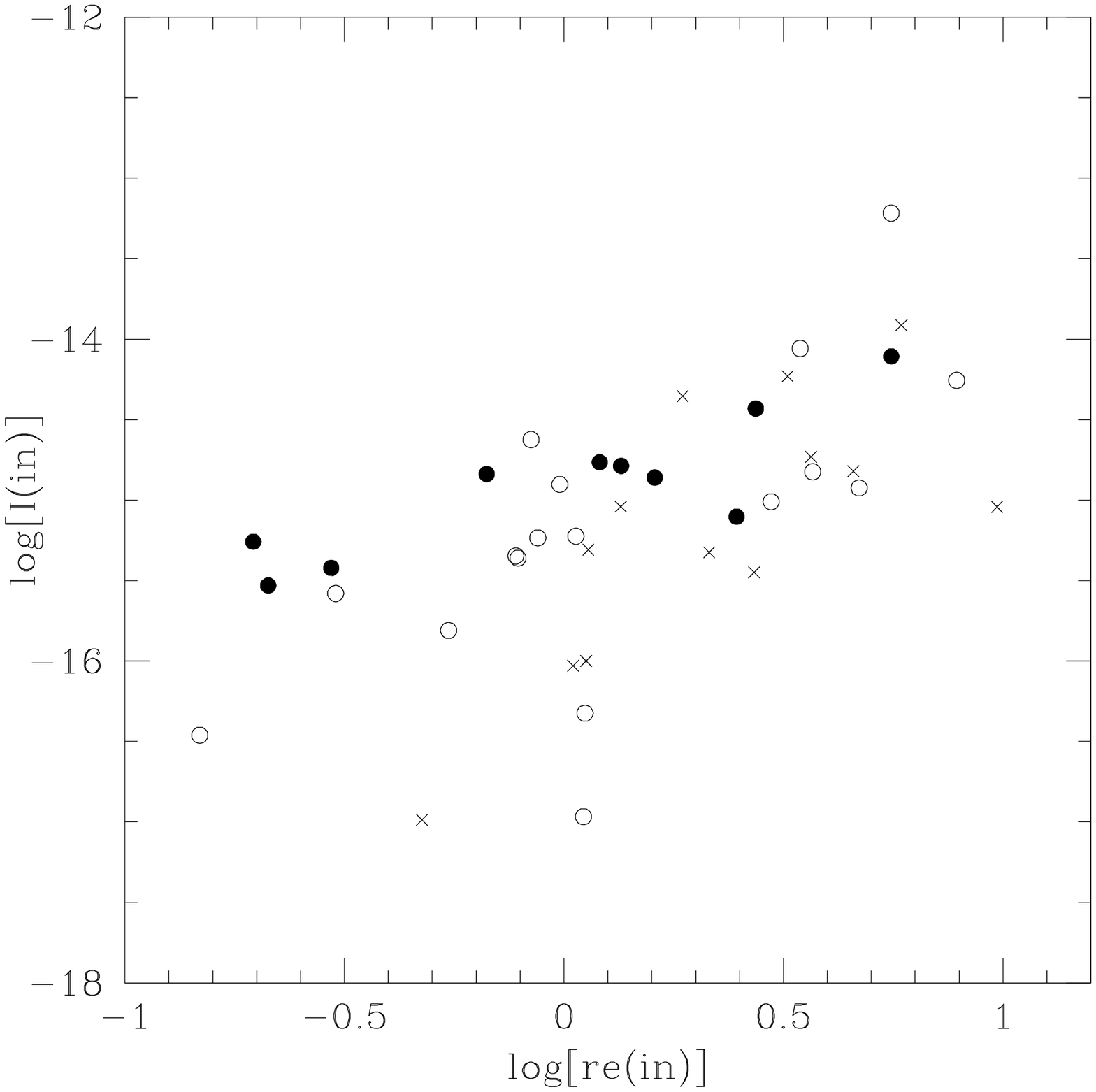}
\caption{The Fundamental Plane equations using half-light radii $r_{e}$ and surface brightnesses $\mu_{e}$ for the inner fitted component (i.e. derived from the parameters $\mu_{in},h_{in}$). $I_{in}$ is the inner component integrated luminosity. Symbols are as in Figure 3.\label{f10}}
\end{figure}

Although the fitting procedure might contribute to producing such a 
correlation (\eg the errors in the fitted parameters might be correlated or 
the profile fitting might somehow conserve the total bulge flux), this is very
unlikely. A similar correlation was noticed in a number of previous studies 
while applying {\em different} fitting techniques: \cite{kormendy77a} found 
that for the bulges of early type galaxies there is a suggestive correlation 
between both parameters ($\mu_{e},r_{e}$) and total bulge luminosity. 
\cite{kent85} found a similar correlation of constant luminosity for the 
{\em disk} parameters (the same correlation for the bulge seemed looser), the
bulges of late type galaxies deviating towards lower luminosities 
(which was also noted by \cite{andredakis95}). 
\cite{jong96c} confirmed that the distribution of points in the {\em disk} 
($\mu,h$) plane shows an upper limit that partly follows the line of constant 
total disk luminosity, while found no correlation in the bulge 
($\mu_{e},r_{e}$) plane. All these authors have used different fitting methods
and profile models but reached similar correlations. 
In fact, the correlation between $\mu_{in},h_{in}$ shown here is equivalent to
the Fundamental Plane relations, shown in Figure~\ref{f10} for our samples,
using the half-light radius $r_{e}$ and surface brightness $\mu_{e}$, derived
from the parameters $\mu_{in},h_{in}$ as described in Section 2.  

(ii) There is a striking correlation between the bulge scale lengths $h_{in}$ 
and exponents $n_{in}$ (middle left panel). The solid curves in this plot 
represent lines of constant total luminosity (the same median value as above) 
for a range of central surface brightnesses. Similar curves for larger total 
luminosities would move to the right of the plot. The dotted line indicates 
the index corresponding to a De Vaucouleurs law ({\em n}=0.25) and the dashed 
line the simple exponential ({\em n}=0) case. The observed correlation for the
bulge parameters follows the curve corresponding to 
$\mu_{in}$=15 mag arcsec$^{-2}$ for scale lengths larger than $\sim$0.03 kpc. 
Then, the points move towards larger central surface brightnesses and 
segregate around $n_{in}$=0.25. This is similar to the result discussed above,
that is, that steeper light profiles are needed to accommodate highest 
central surface brightness bulges. There is significantly larger scatter
in the equivalent ($n_{out}$,$h_{out}$) diagram for the disk parameters 
(middle right panel), but most of the points follow here too a curve of 
constant total luminosity that corresponds to ($R$ band) $\mu_{out}\approx$20
mag arcsec$^{-2}$ (for the assumed median). 

(iii) A correlation between the bulge exponent $n_{in}$ and central surface 
brightness $\mu_{in}$ is implied by correlations (i) and (ii) and is shown 
in the lower left panel of Figure~\ref{f9}. The scatter increases 
significantly for $n_{in}\le$0.4, the points mostly segregating around 
$n_{in}$=0.25 (De Vaucouleurs law). The solid curves indicate constant total 
luminosity (the same median value as in the previous plots) for distinct 
values of the bulge scale lengths (here again, curves of larger total 
luminosities move to the right). This plot shows a limited combination of 
bulge profile shapes and central surface brightnesses, that is, there are no 
bulges with high $\mu_{in}$ and flat profiles or with low $\mu_{in}$ and steep
profiles. For the 
equivalent disk parameters $n_{out}$, $\mu_{out}$ (plotted against each other 
on the lower right panel) there is large scatter and no significant 
correlation, the points segregating within a relatively narrow range of scale 
lengths.

Many of the correlations found here were previously noticed for various
galaxy samples. In early type galaxies (E, S0 and bright dwarfs), the 
parameters describing a generalized exponential were found to scale with 
bulge luminosity, in the sense that brighter galaxies show smaller {\em n} and
scale lengths {\em h}  and higher central surface brightness $\mu$ 
(\cite{caon93,jerjen97}). It was suggested that these relations reflect the 
dependence of the total light distribution on the underlying total galaxian 
mass \ie the depth of the gravitational potential 
(\cite{young94,andredakis95}). Similar {\em n}-luminosity and 
{\em n-h} relations were shown to also hold for the bulges of spiral galaxies 
(\cite{andredakis95,habib99}), adding evidence for a similarity between early spiral 
bulges and (intermediate to low luminosity) ellipticals in terms of projected 
light distribution (kinematic similarities between the two classes are already
known). Finally, good correlations between effective bulge radius and surface 
brightness or total bulge luminosity are found for bulges of spirals up to Sb 
types, these being independent of {\em n} value, whereas later spiral types 
were found to systematically deviate from these relations towards lower 
luminosities, this might be indicating a different formation process for these
objects (see also \cite{balcells94}).
Within the limited range of bulge luminosities (see above) we find a
correlation between $\mu_{e}$ and bulge luminosity only for the Warm Seyfert 
1 subsample, while no such correlation is seen for either the Warm Seyfert 2
or the Cold samples. This again argues in favour of intrinsically different
bulge properties between the two Seyfert types. 

Considering next the disk component, \cite{vitores96b} find a good correlation
between disk scale length and total galaxy luminosity for their sample of 
emission line galaxies, indicating that these are disk-dominated systems. We 
searched for correlations between each of the three parameters characterizing 
the disk component and its total luminosity. There is a large range in the 
characteristic parameters for a limited range of total luminosities (as seen
also from Figure~\ref{f9}), but we find some correlation between 
$\mu_{out}$ and $L_{out}$, in particular for Seyfert 2s. We find no 
correlation between any of the disk parameters and the total (broad-band) 
galaxy luminosities, which is not surprising given the significant 
contribution of nuclear light to the total luminosities of Seyfert galaxies 
(in both types) and of additional components such as tidal features to the 
total luminosities of Cold galaxies (these features were excluded when fitting
the disk profiles).

It is interesting to also consider any possible inter-correlations between the
equivalent bulge and disk parameters. \cite{jong96c} found that using an exponential 
bulge ({\em n}=1) an important correlation between $r_{e}$ and $h$ becomes 
apparent, not seen when fitting a De Vaucouleurs profile to the bulge. A 
similar correlation between bulge and disk scale lengths was noticed by
\cite{habib99} for early type disk galaxies.  Such
correlations could indicate coupling between the bulge and disk formation in 
spiral galaxies. We find no significant correlations when using the 
$\mu_{in},h_{in}$ bulge parameters. However, the half-light equivalent 
parameters, $\mu_{e},r_{e}$ appear to correlate with $\mu_{out},h_{out}$
for the Cold sample galaxies only. The coupling between bulge and disk 
confirms our earlier results (Section 3.2.2) for these objects.

\section{Conclusions}

The main conclusions drawn from the parametrization of the light profiles
for our Warm and Cold samples, are as follows: 

1. Through different indicators such as morphological classification, 
Bulge/Disk ratio (from light profile fitting) and relative predominance of the
inner component (from concentration indices and aperture photometry) we
confirm previous results suggesting that Seyfert 1 nuclei tend to reside in 
earlier type hosts compared to Seyfert 2s. The IR-Cold galaxies, on the other 
hand, show complex morphologies and tidal features and are predominantly 
disk-dominated systems.

2. Most concentration indices fail as indicators of morphological type, due to
their sensitivity to the presence of additional features such as bars, rings 
or tidal extensions. The half-light radius $r_{1/2}$ on the other hand is 
correlated with host type T and galaxy size. For all our samples we find a 
correlation between host sizes (larger) and morphological type (later). 
 
3. The Warm Seyfert bulges have similar shapes and their parameters span 
similar ranges for type 1 and 2 nuclei. However, the former tend to have 
larger central surface brightness ($\mu_{in}$) and smaller median scale 
lengths ($h_{in}$). The IR-Cold galaxies have bulges with fainter $\mu_{in}$, 
larger $h_{in}$ and flatter shapes (larger $n_{in}$) compared to the IR-Warm 
sample. Finally, the Warm Seyfert bulges have overall brighter, while Cold 
bulges fainter, $\mu_{in}$ compared to normal spirals or optically-selected 
emission line galaxies with similar morphologies.

4. There is a good overlap in the disk properties of our different samples, 
that is, independently of IR colours and nuclear activity stage. The light 
profiles have similar shapes, best fit with $n_{out}\ge$0.6 and the disk scale
lengths are similar to those of normal spirals (up to Sbc types). The Warm 
Seyferts (in particular type 2s) tend to have brighter blue disk surface
brightness $\mu_{Bout}$ than normal galaxies, while Cold disks are similar to 
normal galaxy disks. The large range in $\mu_{Bout}$ found for our samples 
argues against a preferred value for this parameter.

5. The complex structure seen for most of our objects (IR Warm and Cold) and 
the nuclear contamination, particularly in Seyfert 1s, induce a large scatter 
in the morphological classification and light concentration indices, thus 
weakening any existing correlation between them and the bulge or disk 
parameters. We only find a tendency for bulge-dominated Seyfert 1s to have 
disks that are flatter inwards and truncated outwards ($n_{out}>$1) and for 
disk-dominated Seyfert 2s to have flatter ($n_{in}>$0.25) bulges.

6. The bulge parameters $n_{in}$ and $h_{in}$ show a bimodal distribution 
around the values $n_{in}$=0.25 and 1 (that is, the De Vaucouleurs and simple 
exponential cases), the first better representing high and the latter lower 
surface brightness bulges. This correlation between $\mu_{in}$ and $n_{in}$ 
indicates a limited combination possible for bulge profile shapes and surface 
brightnesses: there are no bulges with high $\mu_{in}$ and flat shapes or low 
$\mu_{in}$ and steep light profiles. However, these two parameters alone
cannot characterize the {\em total} bulge luminosity, bulges having a variety 
of scale lengths. We also find that the bulge total luminosities span a narrow
range for a variety of ($\mu_{in},h_{in}$) or ($n_{in},h_{in}$) combinations. 

7. Disks can have a variety of profile shapes for similar central surface 
brightness and a limited range of scale lengths.
Any correlation between the disk parameters is subject to large 
scatter, due to the presence of additional components that affect the light 
profile, but we still find a limited range of total disk luminosities that can
be characterized by the ($h_{out},n_{out}$) parameters; for a variety of 
combinations, points tend to cluster around a constant $\mu_{out}$.

8. A last point concerns additional components (bars, rings, spiral or 
tidal features) that contribute light in excess of the bulge and disk 
components. In \cite{thesis} we have quantified these, through the definition 
of an ``excess'' index and find that this effect is more important for late 
morphological types and for the more disturbed Warm Seyfert 2 and Cold 
galaxies. 

In the present Paper III, we have shown that the host galaxies of Warm Seyfert
types 1 and 2 show significantly different light distributions, at large 
enough radii (outside the central 2 kpc) that cannnot be attributed to any
nuclear obscuration effects. Seyfert 1 nuclei tend to reside in bulge 
dominated hosts, with steeper light profiles (larger light concentration) and
smaller sizes (see also Paper II). In Seyfert 2 hosts the light is less 
centrally concentrated and the bulge component less prominent. In addition,
Seyfert 2 hosts tend to have more disturbed morphologies and overall larger
sizes. These differences are suggestive of an evolutionary connection between
the two Seyfert types, that might be related to recent interactions/mergers.
This suggestion is consistent with our results in Paper II, indicating
larger dust content and disk star formation in Seyfert 2 galaxies.
In this respect, it is intriguing that the Warm Seyfert 2 properties are 
intermediate between those of Warm Seyfert 1s and Cold galaxies. The latter
reside in late-type hosts of strongly interacting systems, have shallower 
light profiles and, as their optical and IR colours indicate, are probably
dominated by strong disk star formation (Paper II). In Paper V we will further
explore the link between the interaction characteristics and the optical 
properties of our samples and will present evidence for a possible 
evolutionary scenario. Before this, in Paper IV, we will explore the colour 
distributions in all our objects and the implications for their stellar 
content.

\acknowledgments
I am grateful to my thesis advisors George Miley and Walter Jaffe for providing
me with stimulation and support throughout the completion of this project.
This research has made use of the NASA/IPAC Extragalactic Database (NED) 
which is operated by the Jet Propulsion Laboratory, California Institute of 
Technology, under contract with the National Aeronautics and Space 
Administration. Part of this work was completed while the author held a 
National Research Council - NASA GSFC Research Associateship.

%
%

\begin{deluxetable}{lrrrr}
\tablecolumns{5}
\tablewidth{0pc}
\tablecaption{Median Fitted Quantities and Errors. \label{tab4}}
\tablehead{
\colhead{Quantity} & \multicolumn{3}{c}{Median} & \colhead{$\sigma_{med}$} \\
\cline{1-5} \\
\colhead{} & \colhead{Seyf 1} & \colhead{Seyf 2} & \colhead{Cold} & \colhead{} \\
}
\startdata
	$\mu_{0in}$ (\(\frac{mag}{arcsec^2}\))  & 8.52 & 12.462& 17.23 & 0.67 \nl
	$h_{in}^{*}$ (kpc) & 3.5E-4 & 0.03 & 0.58 & 28\% \nl
	$n_{in}$     & 0.26 & 0.30 & 0.66 & 0.05 \nl
	$\mu_{0out}$ (\(\frac{mag}{arcsec^2}\)) & 21.12 & 20.08 & 21.18 & 1.20 \nl
	$h_{out}^{*}$ (kpc) & 5.43 & 5.47 & 6.14 & 25\% \nl
	$n_{out}$    & 1.25 & 1.35 & 1.07 & 0.28 \nl
	$\epsilon$   & 0.17 & 0.37 & 0.42 & 0.01 \nl
	$C_{I/O}$    & 2.14 & 0.97 & 0.78 & \nodata \nl
	$N/D$        & 1.36 & 0.39 & 0.24 & \nodata \nl
	$N/T$        & 0.58 & 0.28 & 0.19 & \nodata \nl
	$C_{31}$     & 5.14 & 4.33 & 3.60 & \nodata \nl
	$r_{1/2}^{*}$ (kpc) & 2.65 & 4.30 & 5.30 & \nodata \nl
\enddata
\end{deluxetable}

\clearpage

%
%

%
%


\clearpage

\clearpage

\clearpage

\clearpage

\figurenum{A1}
\figcaption[fApp1.ps]{Warm Seyfert 1: Isolated object with peculiar morphology: elongated nucleus in E-W direction and a tidal-like feature to the SW. Blue nucleus and an horizontal red stripe extending to the E. On the SW a very blue region gives rise to the blue bump at $\sim$3 kpc in the colour profiles. The red bump at 6 kpc corresponds to a red feature on the W, probably due to dust extinction. $(B-R)$ colour (left)and H$\alpha$ emission (right) maps with size 21$\times$21 kpc. The H$\alpha$ emission is centered on the continuum nucleus (black contours) and appears to be stronger on the E. \label{fApp1}}

\figurenum{A2}
\figcaption[fApp2.ps]{Warm Seyfert 2: Member of a system of strongly interacting galaxies, exhibiting large tidal tails. Within $\sim$2-8 kpc the inner disk and a ring-like structure appear as excess light in the surface brightness profiles. At larger radii the knotty spiral arms dominate the light from the object. At $\sim$ 12 kpc a large tidal arm appears on the W side. The red ring structure on the colour map is due to dust, particularly affecting the $B$ image. (The diffuse redder, than the galaxy disk, on the NE end of the colour map is the closest companion galaxy, while the very red spot in the same region is an overlapping star). The spiral arms are composed by bright knots of emission, probably star forming regions, as the H$\alpha$ image is indicating. $(B-R)$ (left) and H$\alpha$ (right) maps with size 39.5$\times$39.5 kpc. \label{fApp2}}

\figurenum{A3}
\figcaption[fApp3.ps]{Cold galaxy: A galaxy with two central knots and complex spiral morphology. The two knots are separated by $\sim$1.8 kpc: the southern is redder, embedded in a bar-like structure elongated E-W, and is more likely to be the galactic nucleus, while the northern is diffuse and very blue, probably a giant HII region. The ellipse fits were centered on the S knot, consequently the bump at 1.8 kpc on the surface brightness profiles and blue dip in the colour profiles, are the imprint of the N knot. A bright knotty spiral arm $\sim$4.5 kpc to the S, is responsible for the jump on the surface brightness profiles and the blue dip in the colour profiles, in this region. The extended multiple spiral structure appears asymmetric and is composed by bright emission knots. The excess light at $\sim$12 kpc in the surface brightness profiles is due to these outer spiral arms, the southern being also the bluer. The H$\alpha$ emission line image shows strong emission associated with the N knot and the 4.5kpc S spiral arm. $(B-R)$ colour (left) and H$\alpha$ (right) maps with size 24.6$\times$24.6 kpc. (This object is studied in detail by Chatzichristou \etal, 1998). \label{fApp3}}


\begin{thebibliography}{}


\bibitem [Adams 1977]{adams77} Adams, T.F. 1977, \apjs, 33, 19

\bibitem [Allen \& Shu 1979]{allen79} Allen, R.J., Shu, F.H. 1979, \apj, 227, 67

\bibitem [Andredakis etal.\ 1996 ]{andredakis96} Andredakis, Y.C., Peletier, R.F., Balcells, M. 1996, ASP Conf.Ser., 91, 86

\bibitem [Andredakis etal.\ 1995]{andredakis95} Andredakis, Y.C., Peletier, R.F., Balcells, M. 1995, \mnras, 275, 874

\bibitem [Andredakis \& Sanders 1994]{andredakis94} Andredakis, Y.C., Sanders, R.H. 1994, \mnras, 267, 283

\bibitem [Balcells \& Peletier 1994]{balcells94} Balcells, M., Peletier, R.F. 1994, \aj, 107, 135

\bibitem [Boroson 1981]{boroson81} Boroson, T. 1981, \apjs, 46, 177

\bibitem [Burstein 1979]{burstein79} Burstein, D. 1979, \apjs, 41, 435

\bibitem [Caon etal.\ 1993] {caon93} Caon, N., Capaccioli, M., D'Onofrio, M. 1993, \mnras, 265, 1013

\bibitem [Chatzichristou 2000b]{paper2} Chatzichristou, E.T. 2000b, \apj, {\em submitted} (Paper II)

\bibitem [Chatzichristou 2000a]{paper1} Chatzichristou, E.T. 2000a, \apj, {\em submitted} (Paper I)

\bibitem [Chatzichristou 1999]{thesis} Chatzichristou, E.T. 1999, PhD Thesis, Leiden Observatory

\bibitem [Dahari 1984]{dahari84} Dahari, O. 1984, PhD Thesis, University of California, Santa Cruz

\bibitem [Davies 1990]{davies90} Davies, J.J. 1990, \mnras, 244, 8

\bibitem [De Grijp etal.\ 1992]{grijp92} De Grijp, M.H.K., Keel, W.C., Miley, G.K., Goudfrooij, P., Lub, J. 1992, \aaps, 96, 389

\bibitem [De Grijp etal.\ 1987]{grijp87} De Grijp, M.H.K., Miley, G.K., Lub, J. 1987, \aaps, 70, 95

\bibitem [De Jong 1996c]{jong96c} De Jong, R.S. 1996c, \aap, 313, 377

\bibitem [De Jong 1996b]{jong96b} De Jong, R.S. 1996b, \aap, 313, 45

\bibitem [De Jong 1996a]{jong96a} De Jong, R.S. 1996a, \aaps, 118, 557

\bibitem [D'Onofrio etal.\ 1994]{onofrio94} D'Onofrio, M., Capaccioli, M., Caon, N. 1994, \mnras, 271, 523

\bibitem [De Vaucouleurs etal.\ 1991]{vaucoul91} De Vaucouleurs, G., De Vaucouleurs, A., Corwin, A., Buta, R.J., \etal 1991, Third Reference Catalog of Bright Galaxies, Springer-Verlag, New York 

\bibitem [De Vaucouleurs 1977]{vaucoul77} De Vaucouleurs, G. 1977, in {\em Evolution of Galaxies and Stellar Populations}, R.B. Larson, Tinsley B.M., Eds., New Haven: Yale Univ. Obs., p.43

\bibitem [De Vaucouleurs 1974]{vaucoul74} De Vaucouleurs, G. 1974, in {\em The Formation and Dynamics of Galaxies}, IAU Symp. {\bf 58}, J.R. Shakeshaft Eds., Reidel, Dordrecht, p.1

\bibitem [De Vaucouleurs \& Freeman 1970]{vaucoul70} De Vaucouleurs, G., Freeman, K.C. 1970, Vistas in Astronomy, 14, 163

\bibitem [Disney etal.\ 1989]{disney89} Disney, M., Davis, J.I., Phillips, S. 1989, \mnras, 239, 939

\bibitem [Disney 1976]{disney76} Disney, M.J. 1976, \nat, 263, 573

\bibitem [Doi etal.\ 1993]{doi93} Doi, M., Fukugita, M., Okamura, S. 1993, \mnras, 264, 832

\bibitem [Franx etal.\ 1989]{franx89} Franx, M., Illingworth, G., Heckman, T. 1989, \aj, 98, 538

\bibitem[Freeman 1977] {freeman77} Freeman, K.C. 1977 in {\em Structure and 
properties of nearby galaxies} IAU 77, eds. Berkhuijsen E.M. and Wielebinski
R., Reidel, Dordrecht, p.3

\bibitem [Freeman 1970]{freeman70} Freeman, K.C. 1970, \apj, 160, 811

\bibitem [Jansen etal.\ 1994]{jansen94} Jansen, R.A., Knapen, J.H., Beckman, J.E., Peletier, R.F., Hes, R. 1994, \mnras, 270, 373

\bibitem [Jedrzejewski 1987]{jedrzejewski87} Jedrzejewski, R.I. 1987, \mnras, 226, 747

\bibitem [Jerjen \& Binggeli 1997]{jerjen97} Jerjen, H., Binggeli, B. 1997, in {\em The Nature of Elliptical Galaxies}, Proceedings of the Second Stromolo Symposium, Eds. M. Arnaboldi, G,S. Da Costa, P. Saha

\bibitem [Jorgensen etal.\ 1992]{jorgensen92} Jorgensen, I.,Franx, M., Kjaergaard, P. 1992, \aaps, 95, 489

\bibitem [Jura 1980]{jura80} Jura, M. 1980, \apj, 238, 499

\bibitem [Keel 1980]{keel80} Keel, W.C. 1980, \aj, 85, 198

\bibitem [Kent etal.\ 1991]{kent91} Kent, S.M., Dame, T., Fazio, G. 1991, \apj, 378, 131

\bibitem [Kent 1987]{kent87} Kent, S.M. 1987, \aj, 93, 816

\bibitem [Kent 1986]{kent86} Kent, S.M. 1986, \aj, 91, 1301

\bibitem [Kent 1985]{kent85} Kent, S.M. 1985, \apjs, 59, 115

\bibitem [Kormendy \& Richstone 1995]{kormendy95} Kormendy, J., Richstone, D. 1995, \araa, 33, 581

\bibitem [Kormendy etal.\ 1994]{kormendy94} Kormendy, J., Dressler, A., Byun, Y.-I., Faber, S.M., Grillmair, C., \etal 1994, in {\em ESO/OHP Workshop on Dwarf Galaxies}, Meylan G., Prugniel P. Eds., ESO, Garching, p.147

\bibitem [Kormendy 1992]{kormendy92} Kormendy, J. 1992, in ``Galactic Bulges'', IAU Symp. 153, Dejonghe H., Habing H., Eds., Kluwer, Dordrecht, p.209

\bibitem [Kormendy \& Bruzual 1978]{kormendy78} Kormendy, J., Bruzual, G. 1978, \apjl, 223, L63

\bibitem [Kormendy 1977b]{kormendy77b} Kormendy, J. 1997b, \apj, 218, 333

\bibitem [Kormendy 1977a]{kormendy77a}Kormendy, J. 1997a, \apj, 214, 359

\bibitem [Kotilainen \& Ward 1994]{kotilainen94} Kotilainen, J.K., Ward, M.J. 1994, \mnras, 266, 953

\bibitem [Khosroshani etal.\ 1999]{habib99} Khosroshani, Habib, G., Wadadekar, Yogesh, Kembhavi, Ajit 1999, ApJ, {\em submitted}

\bibitem [Lahav etal.\ 1995] {lahav95} Lahav, O., Naim, A., Buta, R.J., Corwin, H.G., De Vaucouleurs, G., \etal 1995, Science, 267, 859

\bibitem [MacKenty 1990]{kenty90} MacKenty, J.W. 1990, \apjs, 72, 231

\bibitem [Martin 1995]{martin95} Martin, P. 1995, \aj, 109, 2428

\bibitem [Morgan 1959]{morgan59} Morgan, W.W. 1959, \pasp, 71, 394 

\bibitem [Morgan 1958]{morgan58} Morgan, W.W. 1958, \pasp, 70, 364 

\bibitem [Okamura etal.\ 1984]{okamura84} Okamura, S., Kodaira, K., Watanabe, M. 1984, \apj, 280, 7

\bibitem [Peletier etal.\ 1994]{peletier94} Peletier, R.F., Valentijn, E.A., Morwood, A.F.M., Freudling, W. 1994, \aaps, 108, 621

\bibitem [Pfenniger \& Norman 1990]{pfenniger90} Pfenniger, D., Norman, C. 1990, \apj, 363, 391

\bibitem [Phillips \& Disney 1993]{phillips93} Phillips, S., Disney, M.J. 1993, \mnras, 203, 55

\bibitem [Prieto etal.\ 1992b]{prieto92b} Prieto, M., Beckman, J.E., Varela, A.M. 1992b, \aap, 257, 85

\bibitem [Prieto etal.\ 1992a]{prieto92a} Prieto, M., Longley, D.P.T., Perez, E., Beckman, J.E., Varela, A.M., Cepa, J. 1992a, \aaps, 93, 557

\bibitem [Romanishin etal.\ 1983]{romanishin83} Romanishin, W., Strom, K.M., Strom, S.E. 1983, \apjs, 53, 105 

\bibitem [Salzer etal.\ 1989]{salzer89} Salzer, J.J., MacAlpine, G.M., Boroson, T.A. 1989, \apjs, 70, 479

\bibitem [Schmidtke etal.\ 1997]{schmidtke97} Schmidtke, P.C., Windhorst, R.A., Mutz, S.B., Pascarelle, S.M., Franklin, B.E. 1997, \aj, 113, 569

\bibitem [Serna 1997]{serna97} Serna, A. 1997, \aap, 318, 741

\bibitem [Sersic 1968]{sersic68} Sersic, J.L. 1968, Atlas de Galaxias Australes, Observatorio Astronomico, Cordoba

\bibitem [Shaw \& Gilmore 1989]{shaw89} Shaw, M.A., Gilmore, G. 1989, \mnras, 237, 903

\bibitem [Simien \& De Vaucouleurs 1986]{simien86} Simien, F., De Vaucouleurs, G. 1986, \apj, 302, 564

\bibitem [Simkin etal.\ 1980]{simkin80} Simkin, S.M., Su, H.J., Schwarz, M.P. 1980, \apj, 237, 404

\bibitem [Valentijn 1990]{valentijn90} Valentijn, E.A. 1990, \nat, 346, 153

\bibitem [Van der Kruit 1987]{kruit87} Van der Kruit, P.C. 1987, \aap, 173, 59

\bibitem [Vitores etal.\ 1996b]{vitores96b} Vitores, A.G., Zamorano, J., Rego, M., Gallego, J., Alonso, O. 1996b, \aaps, 120, 385 

\bibitem [Vitores etal.\ 1996a]{vitores96a} Vitores, A.G., Zamorano, J., Rego, M., Alonso, O., Gallego, J. 1996a, \aaps, 118, 7

\bibitem [Wehinger \& Wyckoff 1977]{wehinger77} Wehinger, P.A., Wyckoff, S. 1977, \mnras, 181, 211

\bibitem [White \& Keel 1992]{white92} White, R.E. III, Keel, W.C. 1992, \nat, 359, 129

\bibitem [Yee 1983]{yee83} Yee, H.K.C. 1983, \apj, 272, 473

\bibitem [Young \& Currie 1994]{young94} Young, C.K., Currie, M.J. 1994, \mnras, 268, L11

\end{thebibliography}
\end{document}